\documentclass[structabstract]{aa}  
\usepackage{graphicx}
\usepackage{txfonts}
\usepackage{natbib}
\bibpunct{(}{)}{;}{a}{}{,}
\newcommand{\NeII}{[Ne\,{\sc ii}]\ }
\newcommand{\NeIII}{[Ne\,{\sc iii}]\ }
\newcommand{\kms}{km\,s$^{-1}$}
\newcommand{\Wmq}{Wm$^{-2}$}
\newcommand{\ergps}{erg\,s$^{-1}$}
\newcommand{\mum}{$\mu$m}
\newcommand{\LNeII}{$L_{\rm{[Ne\,II]}}$}
\newcommand{\LX}{$L_{\rm X}$}

\newcommand{\Msun}{M$_{\odot}$}

\newcommand{\revised}{\rm}

\begin{document}
\title{An outflow origin of the \NeII emission in the T~Tau triplet\thanks{Based on observations performed at ESO's La Silla-Paranal observatory under DDT programme 280.C-5035}}


\author{R.~van~Boekel\inst{1} 
        \and M.~G\"udel\inst{2,1,3} 
        \and Th.~Henning\inst{1} 
        \and F.~Lahuis\inst{3,4}
        \and E.~Pantin\inst{5}
         }

\institute{Max-Planck Institute for Astronomy, K\"onigstuhl 17,
           D-69117 Heidelberg, Germany
           \and
           Institute of Astronomy, ETH Zurich, 8093 Zurich, Switzerland
           \and
           Leiden Observatory, Leiden University, P.O. Box 9513, 2300 RA Leiden, Netherlands
           \and
           SRON Netherlands Institute for Space Research, P.O. Box 800, 9700 AV Groningen, Netherlands
           \and Laboratoire AIM, CEA/DSM-CNRS-Universit\'e Paris Diderot, IRFU/Service d'Astrophysique, B\^at. 709, CEA/Saclay, 91191 Gif-sur-Yvette Cedex, France
}

   \date{Received 1 December 2008 / Accepted 22 January 2009}

  \abstract
{The 12.81\,$\mu$m \NeII line has recently gained interest as a potential tracer of gas in the tenuous surface layers of circumstellar disks and in outflow-related shocks. Evidence has been found for a proportionality between \NeII emission and X-ray luminosity, supporting the hypothesis that X-rays are responsible for the required ionization and heating of the gas. Alternatively, ionization and heating by EUV photons and in J-type (dissociative) shocks has been proposed.}
{The T\,Tau multiple system harbors three stars with circumstellar disks, at least one strong X-ray source (T\,Tau\,N), and regions of shocked gas in the immediate vicinity. ISO and Spitzer spectra revealed remarkably strong \NeII emission, but because of insufficient spatial and spectral resolution those observations could neither pinpoint \emph{where} in the system the \NeII emission arises, nor identify the emission mechanism. We aim to clarify this by observing the system with enough resolution to spatially separate the various components and spectrally resolve the line emission.}
{We performed high-resolution ($R$=$30000$) spectroscopy of the T\,Tau triplet at $\sim$0\farcs4 spatial resolution with VISIR at the VLT early February 2008. We spatially separated T\,Tau\,N from the southern close binary T\,Tau\,S, as well as the structures of shocked gas surrounding the stars. The individual southern components Sa and Sb remained spatially unresolved in our observations.}
{The dominant component of \NeII emission is centered on T\,Tau\,S and has a spatial extent of $FWHM$$\sim$1\farcs1 in a Gaussian fit. We detect spatially extended red-shifted emission NW of the system and fainter blue-shifted emission to the SE, which we associate with the N-S outflow from T\,Tau\,S. Only a small fraction of the \NeII emission appears directly related to the X-ray bright northern component. Shocks may account for a substantial and possibly dominant fraction of the observed \NeII emission.
We estimate the total \NeII flux to be 23$\pm$6\,$\times$10$^{-16}$~\Wmq, in good agreement with the values measured by ISO in late 1997 and Spitzer in early 2004.}
{Our observations show that outflows rather than the disk surface may dominate the observed \NeII emission in stars with strong outflow activity. We propose \NeII emission in jets as a major factor causing the observed large scatter in the \LX \ vs. \LNeII \ relation. We argue that T\,Tau\,S is the driving source of the T\,Tau ``NW-blob''.}

\keywords{
stars: pre-main sequence --
stars: individual: T\,Tau --
circumstellar matter --
infrared: stars --
shock waves --
X-rays: stars
}

   \maketitle

\section{Introduction}
Circumstellar disks are indispensable in the formation process of stars and are the birthplaces of planetary systems. Infrared observations of the refractory material (``dust'') show that the ingredients needed to form terrestrial planets and cores of giant gas planets are present and in place \citep[see][for an overview of dust in proto-planetary disks]{2007prpl.conf..767N}. The bulk of the disk mass is present in gaseous form, and is very difficult to observe due to its very low average opacity.

Infrared emission in the \NeII fine structure transition at 12.81\,\mum \ has recently gained interest as a potential tracer of gas in the tenuous upper layers of circumstellar disks, or of small amounts of gas in debris disks. The intrinsically much weaker \NeIII transition at 15.55\,\mum \ is observed only in exceptional cases \citep{2007ApJ...665..492L}. If Ne$^+$ is present and the gas is heated to $\gtrsim$10$^3$\,K, the fine structure transitions are excited and we may observe \NeII emission \citep{2007ApJ...656..515G, 2007ApJ...663..383P, 2007ApJ...665..492L,2008ApJ...688..398E}. Several candidate mechanisms for the ionization and heating of the gas have been proposed.
Neon can be ionized via K-shell absorption of stellar X-rays, X-ray irradiation also heats the gas to several thousand Kelvin \citep{2007ApJ...656..515G,2008ApJ...688..398E}. Alternatively, EUV photons may ionize neon and absorption of radiation by small grains or PAHs heats the gas via the photoelectric effect \citep{2008ApJ...683..287G}. Strong, dissociative (J-type) shocks constitute a third possible ionization and heating mechanism \citep[e.g.][]{1999A&A...348..877V}.

\cite{2007ApJ...656..515G} model the effect of X-ray irradiation on the surface of a circumstellar disk, demonstrating that an X-ray source with a luminosity typical of young stars provides significant ionization of Neon. The X-ray irradiation also heats the gas in the disk atmosphere to several thousand K out to a radius of $\sim$20\,AU, beyond which there is a rather abrupt drop in gas temperature. Including mechanical heating from e.g. wind-disk interaction only makes a minor quantitative difference. Therefore, the \NeII emission is restricted to the inner $\sim$20\,AU of the disk.

{\revised To clarify the nature of \NeII emission in YSOs there are currently two viable approaches. The first is a ``statistical'' approach, in which \NeII luminosities of a sample of sources are compared to other observables, such as the X-ray luminosity, to search for correlations. Observations with the Spitzer Space Telescope have provided robust measurements of \NeII fluxes of tens of young stars \citep[e.g.][]{2007ApJ...665..492L}. Current space based observations are very sensitive, allowing relatively weak \NeII lines to be detected, but leave the line emission spectrally and spatially unresolved, limiting their use as diagnostics for the emission mechanism. The second approach is to perform detailed observations of individual objects, in which the \NeII emission is spectrally or spatially resolved. Such observations are currently only possible with ground based instrumentation, strongly reducing sensitivity compared to space-based measurements, and are only feasible for comparatively \NeII bright objects. In this paper, we follow the second approach.}

{\revised In a pioneering study following the statistical approach, \cite{2007ApJ...663..383P} suggested a correlation between the strength of the \NeII emission and the X-ray luminosity in young stars surrounded by circumstellar disks}. However, their sample constituted only 4 detections covering 0.2\,dex in \LNeII \ and \LX, leaving the proposed correlation tentative. Based on a somewhat larger sample, \cite{2007ApJ...664L.111E} cast doubt on the proposed relation between \LNeII \ and \LX. They suggest that \LNeII \ may instead be correlated to the accretion rate and propose EUV radiation from accretion shocks, rather than stellar X-rays, to be the main agent for irradiation of the disk surface and formation of the \NeII line. {\revised In a study following the statistical approach that we have conducted in parallel to the work presented here, we have extended the diagram to 33 objects with both \NeII and X-ray detections, spanning $\sim$2\,dex in both \LNeII \ and \LX \ \citep[][ in prep.]{guedel_statistical}}, and do find a general trend of increasing \LNeII \ with increasing \LX. However, large scatter is evident, {\revised casting doubt on the proposed direct relationship between \NeII emission and X-rays, and arguing that the \NeII generation in young stars is more complex. In particular, sources with strong outflow activity are typically found to be over-luminous in the \NeII line.}

\vspace{0.15cm}

{\revised Here we investigate the nature of the \NeII in YSOs by performing a detailed study of the prototype of young, low mass stars: T\,Tau \citep{1945ApJ...102..168J,ambartsumian1947,ambartsumian1949}. The T\,Tau system suits our objectives particularly well for several reasons: 1) it is a multiple system with 3 known disk-bearing young stars within the central arcsecond, of which at least one is a strong X-ray source (T\,Tau\,N); 2) it contains diffuse regions of shocked gas related to outflows on scales of several arcseconds; 3) it shows very strong \NeII emission, allowing ground based studies at high spatial and spectral resolution.

Located in the Taurus molecular cloud at a distance of 148\,pc \citep{2007ApJ...671..546L}, the T\,Tau system consists of the optically visible northern component T\,Tau\,N and the ``infrared companion'' T\,Tau\,S approximately 0\farcs7 to the south \citep{1982ApJ...255L.103D}. The latter is itself a close binary \citep{2000ApJ...531L.147K} with a current projected separation between the components Sa and Sb of $\sim$0\farcs13 or $\sim$19\,AU  \citep[e.g.][]{2008A&A...482..929K}. Mass estimates of the system members are $\sim$2\,\Msun, $\sim$2.1\,\Msun \  and  $\sim$0.8\,\Msun \ for T\,Tau\,N, Sa, and Sb, respectively \citep[e.g.][]{2008A&A...482..929K}. T\,Tau suffers modest and time-variable line of sight extinction of $A_{\rm V}\sim$1\,mag \citep{1974A&AS...15...47K,2005AstL...31..109M,2007ApJ...671..546L}. Towards Sb the extinction is estimated to be $\sim$15\,mag \citep{2005ApJ...628..832D}, the extinction towards Sa is substantially higher and may be several tens of magnitudes \citep[][]{2005ApJ...628..832D}. All three components show strong infrared excess emission, indicative of circumstellar disks \citep{1966ApJ...143.1010M,2008ApJ...676.1082S}. The heliocentric radial velocity of T\,Tau\,N is +19\,\kms \ \citep{1986ApJ...309..275H}, those of Sa and Sb are $+$22 and $+21$~\kms, respectively \citep{2005ApJ...628..832D}.}

{\revised One of the three stars, T\,Tau\,N, is a very bright, luminous, and variable X-ray source, the X-rays being predominantly of coronal origin. \citep{1995A&A...297..391N,2000A&A...356..949S,2007A&A...468..529G}. The other two stars are probably also prominent X-ray emitters but the strong absorption by the high intervening gas columns makes their detection close to the bright T\,Tau\,N very difficult \citep[see Sect.~\ref{sec:emission_mechanism} below;][]{2007A&A...468..529G}. The T\,Tau system was also shown to exhibit strong \NeII emission by \cite{1999A&A...348..877V}, who tentatively associate this emission with shocked gas in the NW~blob.}

\vspace{0.15cm}

{\revised In this study, we seek to clarify the origin of the \NeII emission in the T\,Tau system. All three known stars, as well as the extended regions of shocked gas}, fall within a single spatial resolution element of existing spectra made with \emph{ISO} and \emph{Spitzer}, leaving the origin of the emission and the underlying mechanism undetermined. We tackle the problem using high spatial ($\sim$0\farcs4) and spectral ($R$$\sim$30000) resolution ground-based observations, resolving the various components of the system.

\begin{figure*}[t]
\centerline{
\includegraphics[height=17.5cm,angle=90]{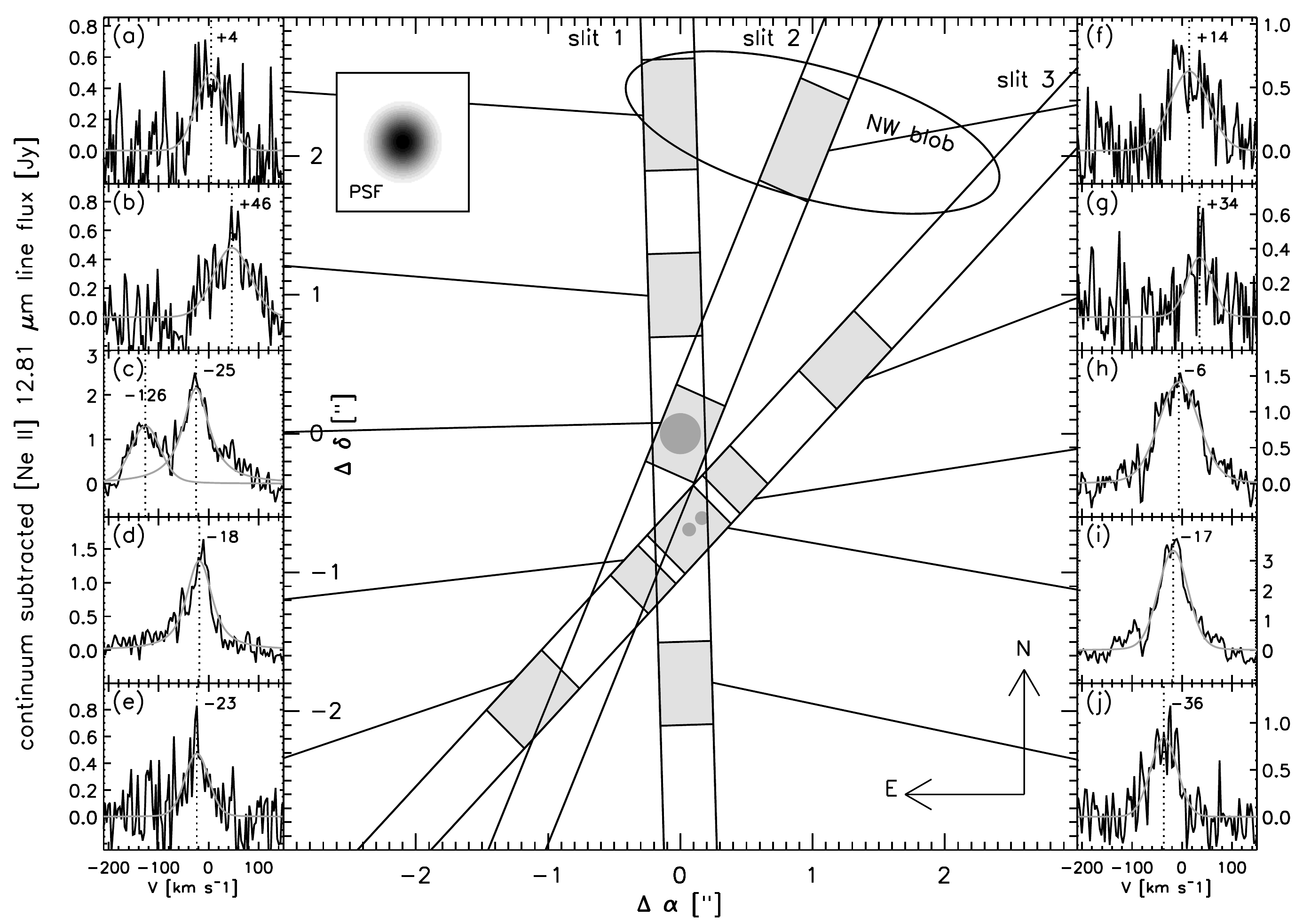}
}
\caption{\label{fig:line_spectra} Continuum subtracted spectra of the T\,Tau system showing the 12.81\,\mum \ \NeII emission line at various positions. The apertures over which the spectra were extracted are shaded grey and solid lines connect the individual spectra and apertures. Approximate radial velocities are determined by fitting Voigt profiles, and are indicated in \kms \ relative to the assumed radial velocity of T\,Tau\,Sa, \citep[$+22$\kms,][]{2005ApJ...628..832D}. Note that the line profiles are generally much broader than the instrumental $FWHM$ of $\approx$\,10\,\kms.}
\end{figure*}

\section{Observations and data reduction}
\label{sec:observations}
T\,Tau was observed with the mid infrared imager and spectrograph \emph{VISIR} \citep{2004Msngr.117...12L}, mounted on \emph{Melipal}, the third of VLTs four 8.2\,m Unit Telescopes. Our primary goal was to obtain high resolution spectra around the \NeII ($^2P_{3/2}-$$^2P_{1/2}$) fine-structure line at 12.81355\,$\mu$m \citep{1985JChPh..83..552Y}. {\revised Spectroscopic observations were performed with 3 slit orientations, each executed during a separate night in February 2008 and calibrated independently (see Table~\ref{tab:log})}. Additionally, imaging in an approximately 0.23\,$\mu$m wide filter centered on 12.81\,$\mu$m was performed, aimed at detecting bright \NeII features at positions not covered by our spectroscopic observations. At a 3$\sigma$ detection threshold of $\sim$5$\times$10$^{-16}$ W\,m$^{-2}$\,arcsec$^{-2}$, we did not detect any such features at radii beyond 1\arcsec \ from the system. Our spectra are much more sensitive to \NeII emission since the high dispersion strongly dilutes the vastly dominant telluric background and continuum dust emission. {\revised Note that the \NeII emission contributes only $\sim$5\% to the system flux integrated over the spectral passband of our imaging filter, the remaining $\sim$95\% being continuum dust emission.} In the remainder of this paper, we will restrict the discussion to the spectroscopic observations. 

In an accompanying paper {\revised (van Boekel et al. 2009, in prep.)} we report on the imaging observations, announcing unexpectedly fast ($\sim$0.25\,mag increase in 4~days) {\revised continuum} brightness variations in T\,Tau\,S at 12.8\,\mum. {\revised We attribute these to fast variations of the irradiation of the disk surface of T\,Tau\,Sa, due to variations in the accretion luminosity of the central source. Observationally, there is no evidence for a direct relation the between the IR continuum variability and the \NeII emission (see also Sect.~\ref{sec:variability}).
}

\subsection{HR spectroscopy around 12.81 $\mu$m}
Long-slit spectroscopy of the T\,Tau system was performed with a slit width of 0\farcs4 and a spectral resolution of $R$$\sim$30000. In order to test various hypotheses to the origin of the \NeII emission we observed the system with 3 different slit orientations, spatially covering the main components of interest (see Fig.\,\ref{fig:line_spectra}). Slit~1 covers the northern and southern component, and is oriented roughly perpendicular to the Sa-Sb separation. Slit~2 covers the northern component and the ``NW blob'', and incidentally catches T\,Tau\,S as well, albeit with significant slit losses. Slit~3 covers the southern component, and is oriented along the Sa-Sb separation. All slits cover some of the diffuse gaseous emission seen in deep AO-assisted near-infrared images \citep[e.g.][]{2007AJ....134..359H,2008ApJ...676..472B,2008A&A...488..235G}.

All observations were performed by ESO staff during the nights starting 2, 3, and 6 February 2008, and are summarized in table\,\ref{tab:log}. A relatively large chop throw of 18\arcsec \ was used to avoid possible faint extended emission to be lost in the sky subtraction procedure. Any sufficiently bright emission within 9\arcsec \ of the central sources would be detectable using our setup. A spectroscopic calibrator was observed along with each science observation for telluric and flux calibration. The atmospheric transmission was calculated using ATRAN \citep{lord92}, a water column of $\approx$2\,mm was found to give a good match to the calibration observations (note that there is only 1 strong water line in the spectral range covered, which is right at the red edge of our spectra and does not interfere with the \NeII emission). Other telluric lines in our spectra are due to O$_3$ and CO$_2$, a strong CO$_2$ line at 12.81224\,$\mu$m is the main spectral feature interfering with the \NeII line. Our telluric correction using ATRAN removes all features to the noise level of our calibration observations (SNR\,$\approx$\,30). Taking the standard deviation in the total system response (including atmospheric transparency) determined from 6 calibration measurements, we estimate the absolute flux calibration of our spectra to be accurate to 5\% (1$\sigma$). {\revised  Radial velocity corrections have been applied to account for the Earth's orbital motion, and the indicated velocities for all spectra shown in this paper are with respect to the systemic velocity of T\,Tau\,Sa \citep[$+$22\,\kms \ heliocentric,][]{2005ApJ...628..832D}.}

VISIR HR spectra suffer from ``fringing'': a modulation of the flux with wavelength, with a peak to peak amplitude of about 15\%. We found this effect to be stable over periods of at least several months by comparing our February data with observations obtained in June 2008. We fitted a spline profile to the calibration observations and divided the science observations by this curve. This procedure yielded satisfactory results, we estimate the amplitude of any possible remaining artifacts to be $\lesssim$2\%.

\begin{table}
\caption{\label{tab:log} Log of the spectroscopic observations with VLT/VISIR.}
\begin{tabular}{lcccr}
target & slit & observing date &  airmass & $T_{int}$ \\
\hline
\\

T-Tau    & slit 1  & 03.02.2008 \ \ 00:34  & 1.42  &    800 \\
HD-28305 & slit 1  & 03.02.2008 \ \ 01:20  & 1.48  &    800 \\
T-Tau    & slit 3  & 04.02.2008 \ \ 00:34  & 1.43  &    960 \\
HD-28305 & slit 3  & 04.02.2008 \ \ 01:26  & 1.51  &    800 \\
HD-28305 & slit 2  & 07.02.2008 \ \ 00:41  & 1.43  &    800 \\
T-Tau    & slit 2  & 07.02.2008 \ \ 01:23  & 1.59  &   1040 \\
\hline
\end{tabular}
\end{table}

\section{Results}
\label{sec:results}

\begin{figure}[t]
\hspace{-0.27cm}
\includegraphics[height=9.35cm,angle=90]{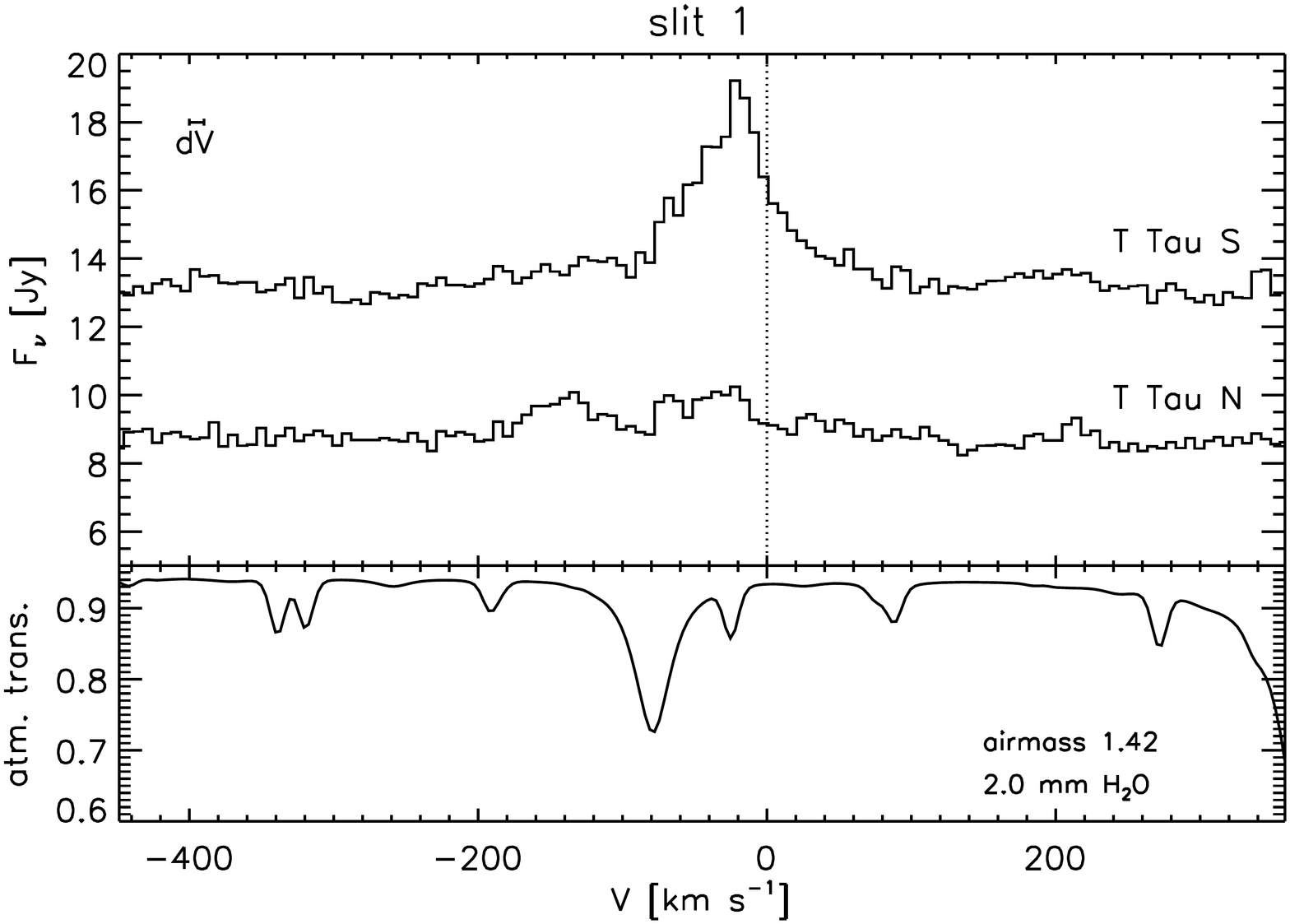}

\caption{\label{fig:spectra1} {\revised High resolution (R$\approx$30000) spectra of the north and south components of the T\,Tau system extracted from slit~1 (see Fig.~\ref{fig:line_spectra}). The spectra shown here were extracted using an approximate method (see Sect.~\ref{sec:results}) and are useful to obtain a general picture of the continuum and \NeII line emission. For the more detailed analysis in this paper we use the continuum subtracted spectra shown in Fig.~\ref{fig:line_spectra}. The velocities are with respect to the radial velocity of T\,Tau\,Sa \citep[$+22$~km\,s$^{-1}$,][]{2005ApJ...628..832D}. The size of a spectral resolution element is plotted in the upper left corner. For reference, the atmospheric transmission is shown in the lower panel. Note that southern binary Sa-Sb remains spatially unresolved in these observations.}}
\end{figure}

In Fig.\,\ref{fig:spectra1} we show the spectra of the northern and southern components, extracted from slit~1 {\revised using the following method. At each wavelength, the profile along the spatial direction is assumed to be the sum of 2 Gaussians (one for T\,Tau\,N and one for T\,Tau\,S), of which the central position, width, and amplitude are fitted to best match the observations. The volume of both Gaussians yields the flux estimates for both sources.} In terms of received energy, both spectra are dominated by continuum emission. The \NeII line is spectrally resolved, and is clearly concentrated on the \emph{southern} component. The extraction of 1-dimensional spectra from our data as shown in Fig.\,\ref{fig:spectra1} provides a useful first impression of the \NeII emission in the system, but discards much of the spatial information in our 2D long-slit spectra.

The continuum radiation we receive is thermal dust emission from the disks around T\,Tau North and South. Its width in the spatial direction ($FWHM$ of Gaussian fits) ranges from 0\farcs43 to 0\farcs48 between the different measurements, compared to 0\farcs40 to 0\farcs49 for the associated calibrators. We thus conclude that the continuum emission at 12.8\,$\mu$m is essentially spatially unresolved for both the northern and southern component {\revised (see also Fig.\,\ref{fig:spatial_profiles} and Sect.\,\ref{sec:qualitative_description})}. At 12.8\,$\mu$m, the dust does not show spectral structure over the small wavelength range covered by our spectra ($\Delta \lambda$$\sim$0.035\,$\mu$m).  {\revised Likewise, the spatial profile is constant over our small spectral range. This allows the continuum emission to be subtracted by fitting the profile in the spectral range where no \NeII emission is detected, and thus to isolate the line emission as well as interpolate of the continuum spatial profile to the wavelengths where we see \NeII emission. We use the continuum subtracted spectra in our analysis.}

\subsection{Qualitative description of the observed \NeII emission}
\label{sec:qualitative_description}
\begin{figure}[t]
\centerline{
\includegraphics[width=9.0cm,angle=0]{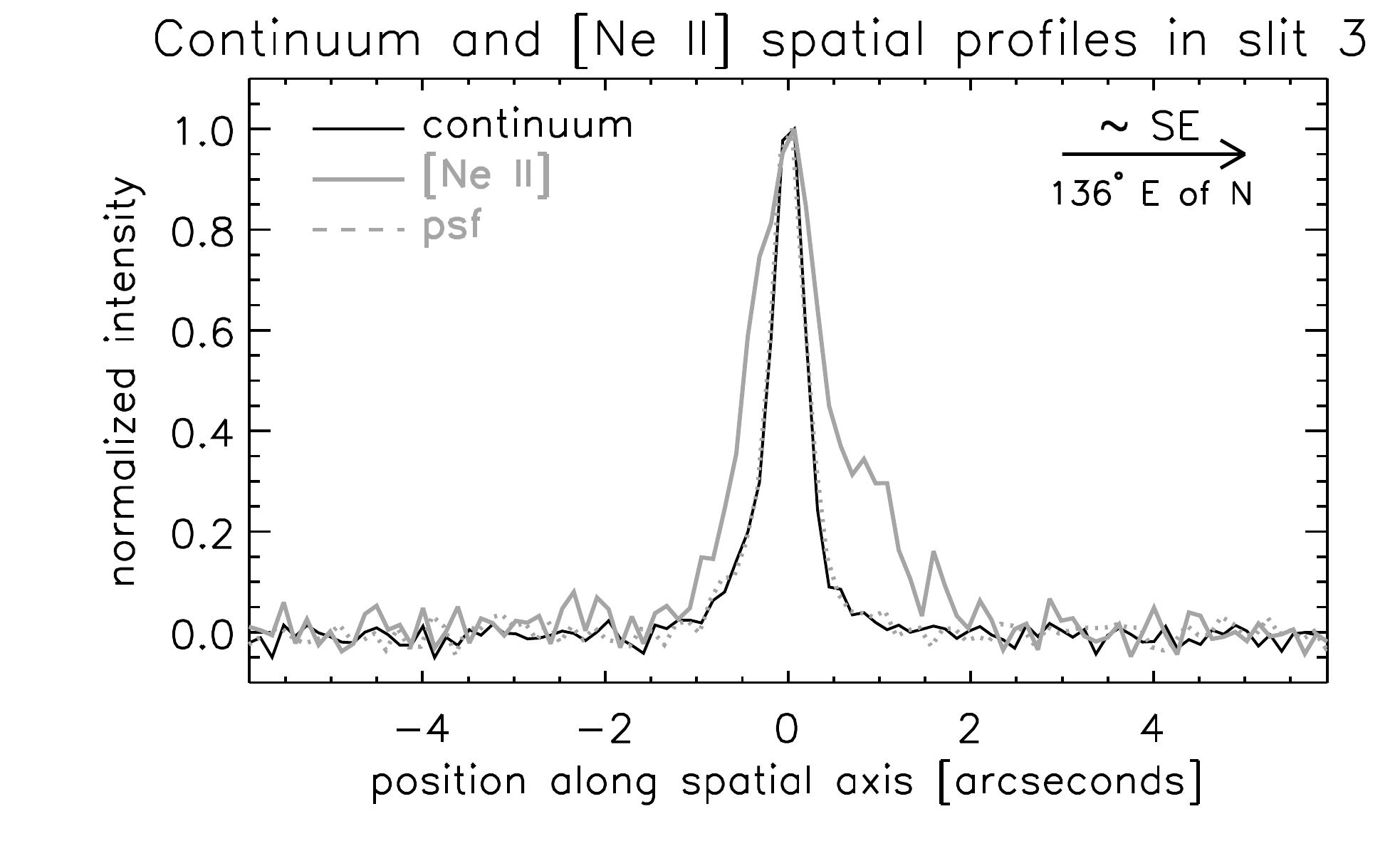}
}
\caption{\label{fig:spatial_profiles} Normalized spatial profiles of the continuum and \NeII emission centered on T\,Tau\,S, as seen through slit~3. The cuts were made at the approximate peak of the \NeII emission at $+$10\,\kms \ with respect to the Sa systemic velocity, and were integrated over 7 pixels along the dispersion direction (corresponding to a passband of 9.8$\times$10$^{-4}$\,\mum \ or 23\,\kms). For reference, the PSF profile extracted from a calibration observation is plotted with a dashed grey curve.}
\end{figure}

Contrary to the continuum dust emission, the observed \NeII emission is clearly spatially extended. {\revised In Fig.~\ref{fig:spatial_profiles} we show spatial profiles of the continuum dust emission and the \NeII emission centered on T\,Tau\,S (after subtraction of the continuum)\footnote{In slit~3, we sampled the continuum blue-ward of the \NeII line between $-$340 and $-$75\,\kms, and red-ward of the line between $+$100 and $+430$\,\kms.} as seen though slit~3, as well as the PSF obtained from the calibration measurement performed immediately after the science observation. The continuum emission remains spatially unresolved, whereas the \NeII emission is clearly spatially resolved. The spatial extent of the line emission is $FWHM$$\sim$1\farcs1 in a Gaussian fit in both slit~1 and slit~3 (N-S and NW-SE direction, respectively).}

{\revised While in Fig.~\ref{fig:line_spectra} we show \NeII spectra extracted in various apertures at key positions in the system, Fig.\,\ref{fig:pv_contours} shows position-velocity diagrams for each of the slits, displaying our data with continuous spatial sampling. The strongest \NeII contribution arises in the 1\farcs1 spatially extended component centered on T\,Tau\,S. This emission is seen in apertures d, i, and h in Fig.~\ref{fig:line_spectra}. } The E-W and NE-SW directions are not covered by our slits and thus the spatial extent in these directions cannot be determined. Note that the size of the emitting region, $FWHM$$\sim$1\farcs1 corresponding to $FWHM$$\sim$160\,AU, is \emph{much larger} than the disk size of either Sa or Sb: due to mutual tidal interactions, either disk cannot be larger than $\sim$5\,AU. The bright component centered on T\,Tau\,S has a velocity centroid that is blue-shifted compared to the stellar radial velocities.

It is possible that the spatially unresolved disks of Sa and Sb contribute to the 1\farcs1 extended component centered on T\,Tau\,S. To estimate this contribution, we made a simple model of the line emission observed in slit~3, consisting of a point source and an extended, Gaussian component. Both components were convolved with the instrumental profile in the spatial direction, taken to be the profile of the unresolved continuum dust emission, and compared to the observed profile. As expected, the extended component contributes dominantly to the total flux, and we estimate the possible contribution of the central point source(s) in T\,Tau\,S to be $\lesssim$0.6\,$\times$10$^{-16}$~W\,m$^{-2}$.

Fainter extended emission is detected at larger distances, out to $\sim$1\farcs9 south of the T\,Tau\,S and $\sim$2\farcs6 north of T\,Tau\,N. The emission in the northern direction is systematically brighter than that in the southern direction. The extended emission seen in the S to SE direction {\revised (apertures e and j in Fig.~\ref{fig:line_spectra})} is blue-shifted, in the N to NW direction we observe red-shifted emission {\revised (apertures b and g)}. We associate this emission with the N-S bipolar outflow discussed by \cite{1994ApJ...430..277B}, i.e. their ``C'' and ``D'' components. The red-shifted ($\approx$$+$40\,\kms) emission in the N direction comes to a halt ($\lesssim$$+$10\,\kms) at the position of the ``North-West blob'' {\revised (apertures a and f)}. This is consistent with the existing notion, that the NW~blob is a bowshock caused by an outflow impinging on ambient material.

At the position of T\,\,Tau\,N {\revised (aperture c)} we find a blue-shifted high-velocity component with a velocity of $\approx$$-$125\,\kms. Both the location and the velocity of this emission match with that of the ``B'' component detected by \cite{1994ApJ...430..277B}, and we associate this emission with a jet from T\,Tau\,N. Additionally, a low-velocity blue-shifted component is seen at the position of T\,Tau\,N. Whether this emission can unambiguously be attributed to T\,Tau\,N is not clear, it may be part of the bright extended component centered on T\,Tau\,S. Interestingly, the total \NeII flux in this component ($\approx$10$^{-16}$\,\Wmq) roughly equals the value predicted by the tentative \LNeII \ vs. \LX \ relation for T\,Tau's very high X-ray luminosity of $\sim$2$\times$10$^{31}$\,\ergps. However, the central velocity of $\approx$$-$22\,\kms with respect to the stellar photosphere of T\,Tau\,N does not agree with a disk surface origin of this emission. If directly related to T\,Tau\,N, this emission may instead arise in a photo-evaporative flow from the T\,Tau\,N disk, of which we only see the approaching part since the receding, red-shifted part is obscured by the disk.

\subsection{The origin of the N-S outflow and the ``NW blob''}
\begin{figure*}[t]
\centerline{
\includegraphics[width=8.0cm,angle=90]{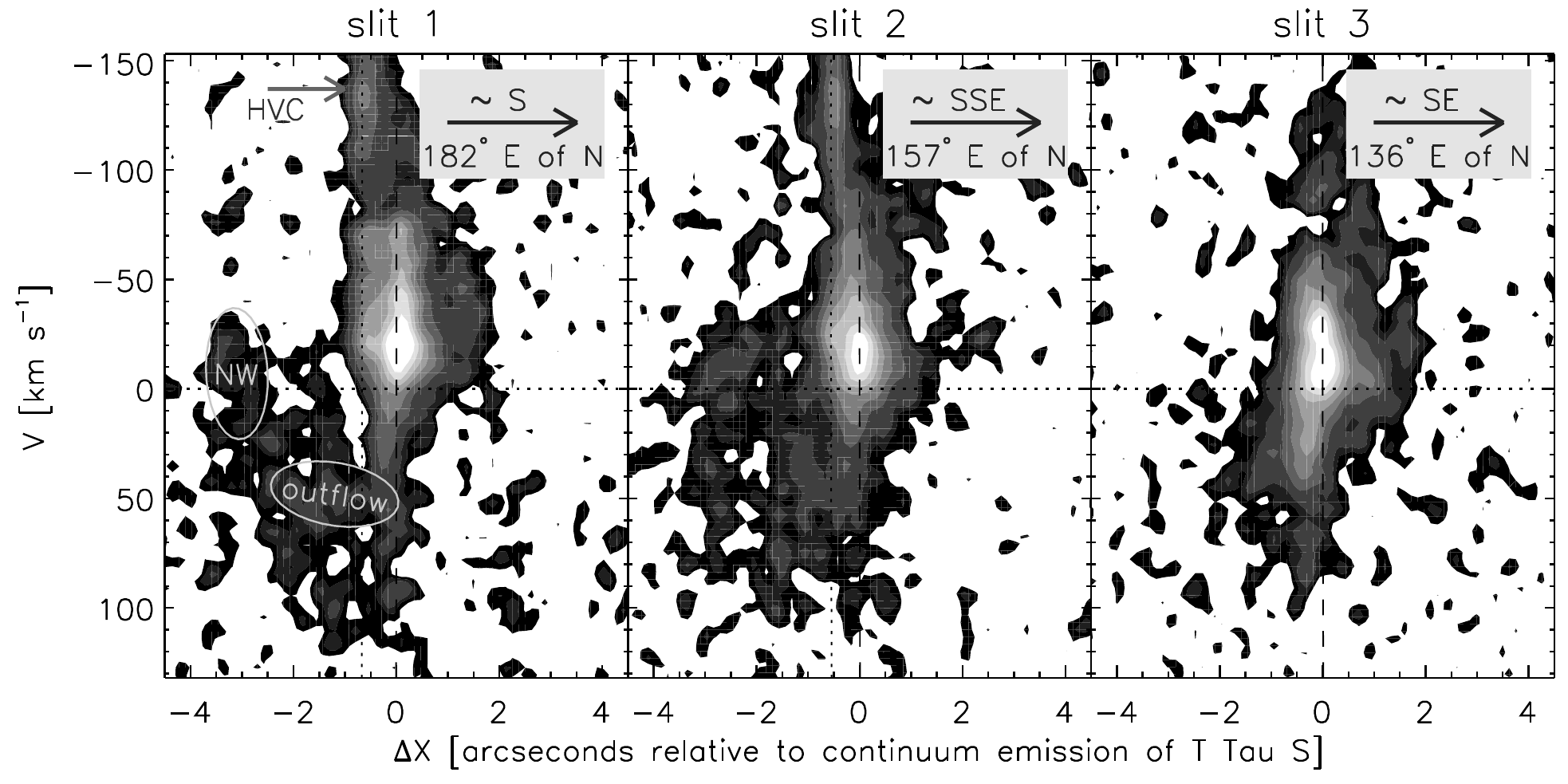}
}
\caption{\label{fig:pv_contours} Position velocity diagrams of the continuum subtracted spectra, showing the \NeII emission in the T\,Tau system along our three slits. The positions of the continuum emission from T\,Tau\,N and T\,Tau\,S are indicated with dashed and dotted vertical lines, respectively. In the left-most diagram (slit~1) we indicated the features we attribute to the northern, red-shifted part of the north-south outflow, and the ``NW~blob''. Contours are drawn at 1, 2, 3, 6, 9, 12, 15, 18, and 21 times the noise level.}
\end{figure*}

Our \NeII position-velocity diagrams (see Fig.~\ref{fig:pv_contours}) clearly show the bright emission centered on T\,Tau\,S, as well as faint, extended emission, particularly to the north. In the $pv$-diagram of slit~1 we indicated the structures we associate with the northern part of the N-S outflow and the ``NW blob''. In both slit~1 and slit~2 we can see that a connected structure is formed by the bright component centered on T\,Tau\,S, material flowing northward with positive radial velocity, and the material in the NW blob that has approximately the stellar radial velocity. This strongly argues for T\,Tau\,S being the source of the N-S outflow, as already proposed by \cite{1994ApJ...430..277B} but later challenged by \cite{1996AJ....111.2403H}. It also shows that the outflow from T\,Tau\,S is indeed the driving force for the NW~blob bowshock.

\subsection{Total \NeII flux recovered}
\label{sec:recovered_flux}
The \NeII line has previously been detected in the T\,Tau system using the ISO satellite in late 1997 \citep{1999A&A...348..877V}, and with the Spitzer Space Telescope in early 2004. The measured line fluxes were 28$\pm$7\,$\times$10$^{-16}$ and 24$\pm$4\,$\times$10$^{-16}$~W\,m$^{-2}$, respectively, during these epochs. This suggests that the \NeII emission is fairly stable on timescales of several years, though measurements at more epochs are required to confirm this. It is also conceivable that short term variations exist, induced by X-ray flares or shocks related to variable accretion in T\,Tau\,S.

With our narrow slit, we do not cover the whole aperture through which the satellite spectra were taken, but we do sample the key regions at least partially. Since our slit width of 0\farcs4 is much smaller than the measured width of $FWHM$=$\sim$1\farcs1 of the \NeII emission centered on T\,Tau\,S, this component suffers from large slit losses. Approximating its spatial intensity profile with a point symmetric Gaussian, we find that a total \NeII  flux from this region of $\sim$11\,$\times$10$^{-16}$~W\,m$^{-2}$. If the actual spatial extent in the E-W direction is smaller than 1\farcs1, we will have over-estimated the contribution from this component.

 North-West of the triplet our slits cover an estimated one third of the region from which the \NeII appears to arise, and we find a flux of $\sim$8\,$\times$10$^{-16}$~W\,m$^{-2}$ (this includes the ``NW blob''). South-East of the triplet we estimate the extended emission to contribute $\sim$2\,$\times$10$^{-16}$~W\,m$^{-2}$, assuming our slits cover one third of the emitting region. About $\sim$2.5\,$\times$10$^{-16}$~W\,m$^{-2}$ of emission is seen at the position of T\,Tau\,N, of which about 40\% stems from the high-velocity blue shifted component that is associated with a jet from T\,Tau\,N, and the rest from the low-velocity blue shifted component.
 
In total, we find a flux of 23$\pm$6\,$\times$10$^{-16}$~W\,m$^{-2}$, where the relatively large error bar reflects the uncertainties induced by the correction for the incomplete spatial coverage of our data (for reference, the total line flux directly observed in the gray shaded apertures indicated in Fig.~\ref{fig:line_spectra} is $\sim$8\,$\times$10$^{-16}$~W\,m$^{-2}$). Thus, we obtain the same line flux as seen by ISO and Spitzer, and show that the \NeII flux of the T\,Tau system was relatively stable over the 1997-2008 period.

\section{Discussion}
Our main goal was to identify the spatial origin of the \NeII emission in the T\,Tau system, with a possible relation between X-rays and \NeII emission in the back of our minds \citep{2007ApJ...656..515G,2007ApJ...663..383P}. The first and most obvious conclusion we can draw is that \emph{in the T\,Tau system there is no obvious direct relation between \NeII emission and X-rays}: whereas T\,Tau\,N is the dominant X-ray source, the \NeII emission arises partly in a bright concentration centered on T\,Tau\,S that may constitute the inner parts of the N-S outflow from this source, and partly in the more extended regions of this outflow as well as in its terminal shock, the NW-blob.

Only a small fraction of the \NeII emission, $\lesssim$10\%, seems directly related to T\,Tau\,N: $\sim$1\,$\times$10$^{-16}$~W\,m$^{-2}$ is emitted in the spatially unresolved high velocity ($-$125\,\kms) component which we associate with the jet of T\,Tau\,N,  another $\sim$1.5\,$\times$10$^{-16}$~W\,m$^{-2}$ is seen at the location of T\,Tau\,N but does not seem to be originating in the disk surface as judged by its radial velocity of $-$22\,\kms with respect to the photosphere of T\,Tau\,N. Instead, it may arise in a photo-evaporative wind from the disk of T\,Tau\,N, or even be an outskirt of the spatially extended 1\farcs1 component centered on T\,Tau\,S.

Thus, while some of the neon emission seen in the T\,Tau system may be X-ray induced, a direct relation between stellar X-rays and \NeII emission from an irradiated disk surface \citep{2007ApJ...656..515G} clearly does \emph{not} hold in the case of T\,Tau. The vast majority of \NeII emission detected in the T\,Tau system does not arise in the surface of any disk.

\subsection{Mechanism responsible for \NeII emission}
\label{sec:emission_mechanism}
What mechanism may be responsible for the ionization and excitation of the neon atoms in the various parts of T\,Tau system? Since the majority of the observed \NeII emission arises in a spatially extended outflow, shocks constitute a primary candidate. In typical jet densities of $\sim$10$^5$\,cm$^{-3}$ the shock velocities required for producing substantial \NeII emission are 70-100\,\kms \citep{1989ApJ...342..306H}. The line of sight velocities we observe in the outflow from T\,Tau\,S are $\sim$40\,\kms, but since the outflow is thought to be oriented close to the plane of the sky the actual velocities could plausibly be 100-300\,\kms. If the jet hits the inner ``wall'' of an outflow cavity under a grazing angle, only the velocity component perpendicular to the wall is relevant for the shock strength and the effective velocities are only a fraction of the jet velocity. All this considered, shocks provide a plausible mechanism for ionizing the material we observe in the \NeII line, but more detailed modeling of the emitting medium is required to be conclusive. The needed temperatures for excitation of the fine structure transition of several thousand Kelvin require shock speeds of 10$-$20\,\kms \ which are easily reached everywhere in the observed outflow.

The second chief candidate mechanism for ionization and heating of the gaseous material is the absorption of stellar X-rays. The effectiveness of this mechanism is difficult to assess. Whereas the level and spectral shape of the X-ray emission from T\,Tau\,N are relatively well known, virtually no X-rays are detected from T\,Tau\,S. However, the high extinction toward both stars in the southern close binary allows for a substantial intrinsic X-ray luminosity of these objects. Sb suffers an estimated extinction of $A_{\rm{V}}$$\sim$15\,mag \citep{2005ApJ...628..832D}; the extinction toward Sa is substantially higher, though its precise value is not well constrained. Adopting $A_{\rm V} = 15$~mag and a conversion to the gas column density appropriate for standard interstellar gas-to-dust mass ratios, $N_{\rm H} = 2\times 10^{21}~A_{\rm V}$~cm$^{-2}$\,mag$^{-1}$ \citep{2003A&A...408..581V}, we simulated the observed X-ray spectrum from T\,Tau\,S assuming that the intrinsic spectrum is identical to the one derived for T\,Tau\,N \citep{2007A&A...468..529G}. We find that the bremsstrahlung portion of the spectrum peaks at 2--3~keV, while radiation at 1~keV is suppressed by 3 orders of magnitude. The total energy flux (in erg~cm$^{-2}$~s$^{-1}$) reaching Earth in the [0.3,10]~keV interval is suppressed by a factor of 4.5. These suppression factors are applicable to possible X-ray emission from Sb, the attenuation of X-rays from Sa would be substantially stronger. {\revised Given the approximate relation between stellar mass and X-ray luminosity for T~Tauri stars \citep{2007A&A...468..425T}, we expect Sb to be intrinsically less luminous in X-rays than T\,Tau\,N by a factor of $\sim$5. The observed X-ray flux from Sb would thus become $\sim$5\% of that of T\,Tau\,N, which agrees remarkably well with the 6.5$\pm$2.7\% that \cite{2007A&A...468..529G} derive for the faint diffuse extention seen in the Chandra image at the approximate position of T\,Tau\,S}. Thus, it is quite possible that the southern stars have significant X-ray emission, and Sa could be as X-ray bright as T\,Tau\,N without being unequivocally detected in the combined X-ray spectrum of the N-S system \citep[see e.g.][who identify the main \emph{observed} X-ray source with T\,Tau\,N]{2007A&A...468..529G}. We note that the similar masses of T\,Tau\,N and Sa \citep[e.g.][]{2008A&A...482..929K} indeed suggest similar X-ray luminosities.

We have made an order of magnitude estimate of the level of \NeII emission that would be induced by X-ray ionization in an extended volume of gas in the outflow from T\,Tau\,S. We do not attempt to model the \NeII emission in detail, but rather wish to establish whether or not stellar X-rays form a plausible ionization mechanism for the outflow material, and focus on the more extended emission at distances of $\gtrsim$1\arcsec. The source of X-rays could be T\,Tau\,N or one of the southern stars. We assume the outflow gas to be hot enough for the 12.81\,\mum \ transition to be excited, and consider only X-ray ionization without additionally requiring heating by X-rays. We followed the methodology of \citet{1983ApJ...267..610K} considering a stellar X-ray source irradiating a gaseous nebula, subject to continuum absorption and ionization. Ionization is to a large extent by photoelectrons produced by the primary photoionization. An average photoionization cross section was used for a gas with cosmic composition ($2.7\times 10^{-22}$~cm$^{-2}$ at 1~keV, \citealt{1999ApJ...518..848I}). For the calculation of the emissivity of the \NeII transition, we followed \citet{2007ApJ...656..515G} using recombination rates from \citet{1982ApJS...48...95S} and \citet{2000asqu.book...95K}. We assuming a characteristic cylindrical outflow volume with a radial depth of 100~AU at a distance of 300-400~AU and a cross-section radius of 150~AU, and a gas density of $10^5$~cm$^{-3}$, {\revised i.e. we mimic the volume of the T\,Tau NW~blob. Furthermore, we take a bremsstrahlung X-ray spectrum approximating that of T\,Tau\,N ($L_{\rm X} = 1.5\times 10^{31}$~erg~s$^{-1}$ in the 0.3--10~keV range, and a coronal electron temperature of $kT = 1.5$~keV, \citealt{2007A&A...468..529G}), an ambient outflow gas temperature of 5000~K, and an ambient ionization fraction of 0.1. We find luminosities of order $10^{29}$~erg~s$^{-1}$ in the \NeII 12.8\,\mum \ line, corresponding to $\sim 4\times 10^{-17}$~W\,m$^{-2}$ of observed flux. In apertures a) and f) in Fig~\ref{fig:line_spectra} we detect a total of $7.6\times 10^{-17}$~W\,m$^{-2}$, which translates to approximately $16\times 10^{-17}$~W\,m$^{-2}$ after correction for the incomplete coverage of NW by our apertures. Thus, the line flux we predict is} somewhat less but within a factor of a few of what the observations show. {\revised Since we know neither the depth of the source nor the density, both of which are extremely important for accurate estimates, our order of magnitude estimate is entirely satisfactory to show that X-ray absorption is a plausible ionization mechanism in the outflow.} We note that this estimate is not based on self-consistent calculations; the production of ambient electrons (for the assumed ionization fraction of the gas) could be due to shock heating although the X-rays themselves help increase the ionization fraction as well.

X-ray heating will be effective relatively close to the stars; extrapolating the effective outer radius of the \NeII emission in the \cite{2007ApJ...656..515G} and \cite{2008ApJ...688..398E} models of $\sim$20\,AU to the X-ray luminosity of T\,Tau\,N, which is approximately 10 times higher than that of the canonical models, we may expect X-ray heating to be effective within $\sqrt{10}$\,$\times$\,20\,AU\,$\lesssim$\,70\,AU. This simple extrapolation ignores optical depth effects, due to the lower densities in the outflow compared to the disk atmosphere in the aforementioned models, X-ray heating may still be effective to somewhat larger distances. However, we detect \NeII emission out to $\sim$2\farcs5 from T\,Tau\,N in the north and south direction, i.e. $\sim$370\,AU in projection. Heating by X-rays will not be effective at these distances, but as argued before shocks provide sufficient heating at every position in the outflow. 

In conclusion, both shocks and X-ray irradiation provide plausible means of ionizing the neon atoms in the outflow from T\,Tau\,S. More detailed investigations are needed to establish which mechanism is dominant. Heating by shocks is probably very effective throughout the outflow, while X-rays are unlikely to contribute significantly to the heating at radii $\gtrsim$100\,AU from the X-ray source(s). A ``hybrid'' model, in which X-rays are responsible for the ionization and shocks provide the heating required to excite the fine-structure transition, may provide the most adequate description of the \NeII emission in the outflow from T\,Tau\,S and those of young stars with strong outflow activity in general.

{\revised
\subsection{Variability of the \NeII emission?}
\label{sec:variability}
The very fast continuum flux variations at 12.8\,\mum \ detected in our VISIR data indicate a variable accretion rate in the central source on scales of a few stellar radii (van Boekel et al. 2009, in prep.), which is likely accompanied by variations in the outflow rate. Such variations may plausibly lead to variations in the \NeII emission of T\,Tau, which is outflow dominated, as we have shown here. However, as outlined in Sect.~\ref{sec:recovered_flux}, we find the \NeII flux to be the same at epochs in 1997, 2004, and 2008, within the substantial uncertainties of the individual measurements. This suggests that the \NeII flux is relatively constant over long periods. The continuum flux of T\,Tau\,S, on the contrary, is known to be strongly variable \citep[e.g.][]{1991AJ....102.2066G}, and in fact it has varied by a factor of 2 between the three aforementioned epochs. Thus, there is no direct relationship between the observed \NeII and IR continuum fluxes. Note that if variations in accretion and outflow rate (as traced by the IR continuum fluxes) would result in changes in the \NeII flux, they would do so with a certain time lag, corresponding to the time needed for the out-flowing material to reach the positions where we observe the \NeII emission. Given the relatively large spatial scales involved, this time lag would be $\sim$1~year to tens of years, depending on the actual (de-projected) outflow velocity and the region under consideration.
}

\section{Summary and conclusions}
We have conducted high spatial (0\farcs4) and spectral ($R$$=$30000) resolution observations of the T\,Tau multiple system, in order to reveal the origin of the strong \NeII emission seen in space based observations in which the emission remained spatially and spectrally unresolved. The T\,Tau system contains 3 circumstellar disks, at least 1 strong X-ray source, and extended regions of shocked gas, thus providing several potential mechanisms producing \NeII emission and an opportunity to test which is effective.

We find that the \NeII emission is \emph{not} concentrated on the X-ray bright northern component, but instead consists of a spatially extended ($FWHM$$\sim$1\farcs1) concentration centered on T\,Tau\,S that is likely the central part of the known N-S outflow from this source, and more diffuse emission associated with this same outflow. The total \NeII flux we derive from our February 2008 observations is, within uncertainties, equal to the flux seen in ISO and Spitzer spectra taken in late 1997 and early 2004, respectively, suggesting that the \NeII emission was not strongly variable over the last decade.

Recently, \cite{2007ApJ...656..515G} proposed that irradiation of the disk surface by stellar X-rays can lead to observable \NeII emission \citep[see also][]{2008ApJ...688..398E}. Indeed, this mechanism was shown to plausibly account for the \NeII emission in several stars \citep{2007ApJ...670..509H,2007ApJ...663..383P}. However, we here show that in the T\,Tau system, there is \emph{no} direct relation between the observed X-ray flux and \NeII emission from a disk surface. {\revised Instead, the \NeII emission arises in an outflow.} While we could not establish the ionization and excitation mechanism of the neon with certainty at every location in the system, {\revised{we show that that both shocks and the absorption of stellar X-rays provide plausible ionization mechanisms for the outflow material. Shocks are the favored mechanism for providing the required heating of outflow material, though X-rays may contribute within $\sim$100\,AU of the stars}.

Generalizing our result, we argue that young stars that exhibit outflows and shocks related to strong accretion activity can show strong \NeII emission that is not directly related to the stellar X-ray emission in the fashion of the \cite{2007ApJ...656..515G} model. We propose outflows to be an important \NeII contributor, that may dominate over the disk surface emission at high outflow rates. This is indeed supported by our statistical study of a large sample of T~Tauri stars that show a correlation between the \NeII luminosity and outflow parameters, and also indicates that stars with jets reveal excessive levels of \NeII emission \citep[][ in prep.]{guedel_statistical}. Lastly, we emphasize the importance of more spatially or spectrally resolved studies of the \NeII line in young stars, such as that of \cite{2007ApJ...670..509H} and this work, in order to search for signatures of outflows and Keplerian rotation in a disk surface. This is a challenging task since in the other stars in which the \NeII line has been detected, it is typically fainter by at least an order of magnitude compared to the line flux observed in the T\,Tau system.

\begin{acknowledgements}
We thank the ESO staff for performing the observations presented in this paper in service mode. We thank Reinhard Mundt, Tom Herbst, Maiken Gustafsson, Rainer K\"ohler, and Thorsten Ratzka for clarifying and inspiring discussions on the circumstellar environment of T\,Tau. MG acknowledges support by the Max-Planck-Institute for Astronomy in Heidelberg, Germany, and Leiden University, The Netherlands, for his Sabbatical stays during which studies of the \NeII radiation partly described here were started. We thank ESO for the opportunity to perform the observations on which this paper is based in Director's Discretionary Time. An anonymous referee is acknowledged for a detailed and constructive review and several recommendations which helped improving the clarity of this paper.
\end{acknowledgements}

\bibliographystyle{aa}
\bibliography{TTau_NeII_vboekel_accepted}

\begin{thebibliography}{42}
\expandafter\ifx\csname natexlab\endcsname\relax\def\natexlab#1{#1}\fi

\bibitem[{{Ambartsumian}(1947)}]{ambartsumian1947}
{Ambartsumian}, V.~A. 1947, Stellar Evolution and Astrophysics (Erevan: Acad.
  Sci. Armenian S. S. R.

\bibitem[{{Ambartsumian}(1949)}]{ambartsumian1949}
{Ambartsumian}, V.~A. 1949, AZh, 26, 3

\bibitem[{{Beck} {et~al.}(2008){Beck}, {McGregor}, {Takami}, \&
  {Pyo}}]{2008ApJ...676..472B}
{Beck}, T.~L., {McGregor}, P.~J., {Takami}, M., \& {Pyo}, T.-S. 2008, \apj,
  676, 472

\bibitem[{{B\"ohm} \& {Solf}(1994)}]{1994ApJ...430..277B}
{B\"ohm}, K.-H. \& {Solf}, J. 1994, \apj, 430, 277

\bibitem[{{Duch{\^e}ne} {et~al.}(2005){Duch{\^e}ne}, {Ghez}, {McCabe}, \&
  {Ceccarelli}}]{2005ApJ...628..832D}
{Duch{\^e}ne}, G., {Ghez}, A.~M., {McCabe}, C., \& {Ceccarelli}, C. 2005, \apj,
  628, 832

\bibitem[{{Dyck} {et~al.}(1982){Dyck}, {Simon}, \&
  {Zuckerman}}]{1982ApJ...255L.103D}
{Dyck}, H.~M., {Simon}, T., \& {Zuckerman}, B. 1982, \apjl, 255, L103

\bibitem[{{Ercolano} {et~al.}(2008){Ercolano}, {Drake}, {Raymond}, \&
  {Clarke}}]{2008ApJ...688..398E}
{Ercolano}, B., {Drake}, J.~J., {Raymond}, J.~C., \& {Clarke}, C.~C. 2008,
  \apj, 688, 398

\bibitem[{{Espaillat} {et~al.}(2007){Espaillat}, {Calvet}, {D'Alessio},
  {Bergin}, {Hartmann}, {Watson}, {Furlan}, {Najita}, {Forrest}, {McClure},
  {Sargent}, {Bohac}, \& {Harrold}}]{2007ApJ...664L.111E}
{Espaillat}, C., {Calvet}, N., {D'Alessio}, P., {et~al.} 2007, \apjl, 664, L111

\bibitem[{{Ghez} {et~al.}(1991){Ghez}, {Neugebauer}, {Gorham}, {Haniff},
  {Kulkarni}, {Matthews}, {Koresko}, \& {Beckwith}}]{1991AJ....102.2066G}
{Ghez}, A.~M., {Neugebauer}, G., {Gorham}, P.~W., {et~al.} 1991, \aj, 102, 2066

\bibitem[{{Glassgold} {et~al.}(2007){Glassgold}, {Najita}, \&
  {Igea}}]{2007ApJ...656..515G}
{Glassgold}, A.~E., {Najita}, J.~R., \& {Igea}, J. 2007, \apj, 656, 515

\bibitem[{{Gorti} \& {Hollenbach}(2008)}]{2008ApJ...683..287G}
{Gorti}, U. \& {Hollenbach}, D. 2008, \apj, 683, 287

\bibitem[{{G{\"u}del} {et~al.}(2009){G{\"u}del}, {Lahuis}, {Henning}, {van
  Boekel}, {some others}, \& {more others}}]{guedel_statistical}
{G{\"u}del}, M., {Lahuis}, F., {Henning}, T., {et~al.} 2009, in prep.,
  http://www.ipac.caltech.edu/spitzer2008/talks/ManuelGuedel.html

\bibitem[{{G{\"u}del} {et~al.}(2007){G{\"u}del}, {Skinner}, {Mel'Nikov},
  {Audard}, {Telleschi}, \& {Briggs}}]{2007A&A...468..529G}
{G{\"u}del}, M., {Skinner}, S.~L., {Mel'Nikov}, S.~Y., {et~al.} 2007, \aap,
  468, 529

\bibitem[{{Gustafsson} {et~al.}(2008){Gustafsson}, {Labadie}, {Herbst}, \&
  {Kasper}}]{2008A&A...488..235G}
{Gustafsson}, M., {Labadie}, L., {Herbst}, T.~M., \& {Kasper}, M. 2008, \aap,
  488, 235

\bibitem[{{Hartmann} {et~al.}(1986){Hartmann}, {Hewett}, {Stahler}, \&
  {Mathieu}}]{1986ApJ...309..275H}
{Hartmann}, L., {Hewett}, R., {Stahler}, S., \& {Mathieu}, R.~D. 1986, \apj,
  309, 275

\bibitem[{{Herbst} {et~al.}(1996){Herbst}, {Beckwith}, {Glindemann},
  {Tacconi-Garman}, {Kroker}, \& {Krabbe}}]{1996AJ....111.2403H}
{Herbst}, T.~M., {Beckwith}, S.~V.~W., {Glindemann}, A., {et~al.} 1996, \aj,
  111, 2403

\bibitem[{{Herbst} {et~al.}(2007){Herbst}, {Hartung}, {Kasper}, {Leinert}, \&
  {Ratzka}}]{2007AJ....134..359H}
{Herbst}, T.~M., {Hartung}, M., {Kasper}, M.~E., {Leinert}, C., \& {Ratzka}, T.
  2007, \aj, 134, 359

\bibitem[{{Herczeg} {et~al.}(2007){Herczeg}, {Najita}, {Hillenbrand}, \&
  {Pascucci}}]{2007ApJ...670..509H}
{Herczeg}, G.~J., {Najita}, J.~R., {Hillenbrand}, L.~A., \& {Pascucci}, I.
  2007, \apj, 670, 509

\bibitem[{{Hollenbach} \& {McKee}(1989)}]{1989ApJ...342..306H}
{Hollenbach}, D. \& {McKee}, C.~F. 1989, \apj, 342, 306

\bibitem[{{Igea} \& {Glassgold}(1999)}]{1999ApJ...518..848I}
{Igea}, J. \& {Glassgold}, A.~E. 1999, \apj, 518, 848

\bibitem[{{Joy}(1945)}]{1945ApJ...102..168J}
{Joy}, A.~H. 1945, \apj, 102, 168

\bibitem[{{Keady} \& {Kilcrease}(2000)}]{2000asqu.book...95K}
{Keady}, J.~J. \& {Kilcrease}, D.~P. 2000, {Radiation} (Allen's Astrophysical
  Quantities), 95--+

\bibitem[{{K{\"o}hler} {et~al.}(2008){K{\"o}hler}, {Ratzka}, {Herbst}, \&
  {Kasper}}]{2008A&A...482..929K}
{K{\"o}hler}, R., {Ratzka}, T., {Herbst}, T.~M., \& {Kasper}, M. 2008, \aap,
  482, 929

\bibitem[{{Koresko}(2000)}]{2000ApJ...531L.147K}
{Koresko}, C.~D. 2000, \apjl, 531, L147

\bibitem[{{Krolik} \& {Kallman}(1983)}]{1983ApJ...267..610K}
{Krolik}, J.~H. \& {Kallman}, T.~R. 1983, \apj, 267, 610

\bibitem[{{Kuhi}(1974)}]{1974A&AS...15...47K}
{Kuhi}, L.~V. 1974, \aaps, 15, 47

\bibitem[{{Lagage} {et~al.}(2004){Lagage}, {Pel}, {Authier}, {Belorgey},
  {Claret}, {Doucet}, {Dubreuil}, {Durand}, {Elswijk}, {Girardot}, {K{\"a}ufl},
  {Kroes}, {Lortholary}, {Lussignol}, {Marchesi}, {Pantin}, {Peletier},
  {Pirard}, {Pragt}, {Rio}, {Schoenmaker}, {Siebenmorgen}, {Silber}, {Smette},
  {Sterzik}, \& {Veyssiere}}]{2004Msngr.117...12L}
{Lagage}, P.~O., {Pel}, J.~W., {Authier}, M., {et~al.} 2004, The Messenger,
  117, 12

\bibitem[{{Lahuis} {et~al.}(2007){Lahuis}, {van Dishoeck}, {Blake}, {Evans},
  {Kessler-Silacci}, \& {Pontoppidan}}]{2007ApJ...665..492L}
{Lahuis}, F., {van Dishoeck}, E.~F., {Blake}, G.~A., {et~al.} 2007, \apj, 665,
  492

\bibitem[{{Loinard} {et~al.}(2007){Loinard}, {Torres}, {Mioduszewski},
  {Rodr{\'{\i}}guez}, {Gonz{\'a}lez-L{\'o}pezlira}, {Lachaume}, {V{\'a}zquez},
  \& {Gonz{\'a}lez}}]{2007ApJ...671..546L}
{Loinard}, L., {Torres}, R.~M., {Mioduszewski}, A.~J., {et~al.} 2007, \apj,
  671, 546

\bibitem[{{Lord}(1992)}]{lord92}
{Lord}, S.~D. 1992, NASA Technical Memorandum 103957

\bibitem[{{Mel'Nikov} \& {Grankin}(2005)}]{2005AstL...31..109M}
{Mel'Nikov}, S.~Y. \& {Grankin}, K.~N. 2005, Astronomy Letters, 31, 109

\bibitem[{{Mendoza V.}(1966)}]{1966ApJ...143.1010M}
{Mendoza V.}, E.~E. 1966, \apj, 143, 1010

\bibitem[{{Natta} {et~al.}(2007){Natta}, {Testi}, {Calvet}, {Henning},
  {Waters}, \& {Wilner}}]{2007prpl.conf..767N}
{Natta}, A., {Testi}, L., {Calvet}, N., {et~al.} 2007, in Protostars and
  Planets V, ed. B.~{Reipurth}, D.~{Jewitt}, \& K.~{Keil}, 767--781

\bibitem[{{Neuh{\"a}user} {et~al.}(1995){Neuh{\"a}user}, {Sterzik}, {Schmitt},
  {Wichmann}, \& {Krautter}}]{1995A&A...297..391N}
{Neuh{\"a}user}, R., {Sterzik}, M.~F., {Schmitt}, J.~H.~M.~M., {Wichmann}, R.,
  \& {Krautter}, J. 1995, \aap, 297, 391

\bibitem[{{Pascucci} {et~al.}(2007){Pascucci}, {Hollenbach}, {Najita},
  {Muzerolle}, {Gorti}, {Herczeg}, {Hillenbrand}, {Kim}, {Carpenter}, {Meyer},
  {Mamajek}, \& {Bouwman}}]{2007ApJ...663..383P}
{Pascucci}, I., {Hollenbach}, D., {Najita}, J., {et~al.} 2007, \apj, 663, 383

\bibitem[{{Shull} \& {van Steenberg}(1982)}]{1982ApJS...48...95S}
{Shull}, J.~M. \& {van Steenberg}, M. 1982, \apjs, 48, 95

\bibitem[{{Skemer} {et~al.}(2008){Skemer}, {Close}, {Hinz}, {Hoffmann},
  {Kenworthy}, \& {Miller}}]{2008ApJ...676.1082S}
{Skemer}, A.~J., {Close}, L.~M., {Hinz}, P.~M., {et~al.} 2008, \apj, 676, 1082

\bibitem[{{Stelzer} {et~al.}(2000){Stelzer}, {Neuh{\"a}user}, \&
  {Hambaryan}}]{2000A&A...356..949S}
{Stelzer}, B., {Neuh{\"a}user}, R., \& {Hambaryan}, V. 2000, \aap, 356, 949

\bibitem[{{Telleschi} {et~al.}(2007){Telleschi}, {G{\"u}del}, {Briggs},
  {Audard}, \& {Palla}}]{2007A&A...468..425T}
{Telleschi}, A., {G{\"u}del}, M., {Briggs}, K.~R., {Audard}, M., \& {Palla}, F.
  2007, \aap, 468, 425

\bibitem[{{van den Ancker} {et~al.}(1999){van den Ancker}, {Wesselius},
  {Tielens}, {van Dishoeck}, \& {Spinoglio}}]{1999A&A...348..877V}
{van den Ancker}, M.~E., {Wesselius}, P.~R., {Tielens}, A.~G.~G.~M., {van
  Dishoeck}, E.~F., \& {Spinoglio}, L. 1999, \aap, 348, 877

\bibitem[{{Vuong} {et~al.}(2003){Vuong}, {Montmerle}, {Grosso}, {Feigelson},
  {Verstraete}, \& {Ozawa}}]{2003A&A...408..581V}
{Vuong}, M.~H., {Montmerle}, T., {Grosso}, N., {et~al.} 2003, \aap, 408, 581

\bibitem[{{Yamada} {et~al.}(1985){Yamada}, {Kanamori}, \&
  {Hirota}}]{1985JChPh..83..552Y}
{Yamada}, C., {Kanamori}, H., \& {Hirota}, E. 1985, \jcp, 83, 552

\end{thebibliography}

\end{document}